\documentclass[12pt, onecolumn, draftcls]{IEEEtran}
\usepackage{epsfig, psfrag, amsmath, amsfonts, amssymb, eucal, endfloat}
\newtheorem{thm}{Theorem}
\newtheorem{cor}[thm]{Corollary}

\newtheorem{prop}{Proposition}

\newcommand{\snr}{\rho}

\nomarkersintext
\renewcommand{\baselinestretch}{1.5}

\DeclareMathAlphabet{\eurm}{U}{eur}{m}{n}
\DeclareMathAlphabet{\mathbsf}{OT1}{cmss}{bx}{n}
\DeclareMathAlphabet{\mathssf}{OT1}{cmss}{m}{sl}
\DeclareMathAlphabet{\mathcsf}{OT1}{cmss}{sbc}{n}

\newcommand{\randomvalue}[1]{\eurm{\uppercase{#1}}}

\DeclareSymbolFont{bsfletters}{OT1}{cmss}{bx}{n}  
\DeclareSymbolFont{ssfletters}{OT1}{cmss}{m}{n}
\DeclareMathSymbol{\bsfGamma}{0}{bsfletters}{'000}
\DeclareMathSymbol{\ssfGamma}{0}{ssfletters}{'000}
\DeclareMathSymbol{\bsfDelta}{0}{bsfletters}{'001}
\DeclareMathSymbol{\ssfDelta}{0}{ssfletters}{'001}
\DeclareMathSymbol{\bsfTheta}{0}{bsfletters}{'002}
\DeclareMathSymbol{\ssfTheta}{0}{ssfletters}{'002}
\DeclareMathSymbol{\bsfLambda}{0}{bsfletters}{'003}
\DeclareMathSymbol{\ssfLambda}{0}{ssfletters}{'003}
\DeclareMathSymbol{\bsfXi}{0}{bsfletters}{'004}
\DeclareMathSymbol{\ssfXi}{0}{ssfletters}{'004}
\DeclareMathSymbol{\bsfPi}{0}{bsfletters}{'005}
\DeclareMathSymbol{\ssfPi}{0}{ssfletters}{'005}
\DeclareMathSymbol{\bsfSigma}{0}{bsfletters}{'006}
\DeclareMathSymbol{\ssfSigma}{0}{ssfletters}{'006}
\DeclareMathSymbol{\bsfUpsilon}{0}{bsfletters}{'007}
\DeclareMathSymbol{\ssfUpsilon}{0}{ssfletters}{'007}
\DeclareMathSymbol{\bsfPhi}{0}{bsfletters}{'010}
\DeclareMathSymbol{\ssfPhi}{0}{ssfletters}{'010}
\DeclareMathSymbol{\bsfPsi}{0}{bsfletters}{'011}
\DeclareMathSymbol{\ssfPsi}{0}{ssfletters}{'011}
\DeclareMathSymbol{\bsfOmega}{0}{bsfletters}{'012}
\DeclareMathSymbol{\ssfOmega}{0}{ssfletters}{'012}

\newcommand{\rvJ}{{\randomvalue{J}}}
\newcommand{\rvU}{{\randomvalue{U}}}

\newcommand{\rva}{{\randomvalue{a}}}	

\newcommand{\rvd}{{\randomvalue{d}}}	

\newcommand{\rvh}{{\randomvalue{h}}}	

\newcommand{\rvm}{{\randomvalue{m}}}	

\newcommand{\rvs}{{\randomvalue{s}}}	

\newcommand{\rvx}{{\randomvalue{x}}}	

\newcommand{\rvy}{{\randomvalue{y}}}	

\newcommand{\rvz}{{\randomvalue{z}}}	

\begin{document}

\title{{Outage-Efficient Downlink Transmission Without Transmit Channel State Information}}
\author{{Wenyi Zhang, {\it Member, IEEE}, Shivaprasad Kotagiri,
{\it Student Member, IEEE}, \\and J. Nicholas Laneman, {\it Senior Member, IEEE}}
\thanks{The work of W. Zhang has been supported in part by NSF through grants NRT ANI-0335302, ITR CCF-0313392,
and OCE-0520324; the work of S. Kotagiri and J. N. Laneman has been supported in part by NSF through grants CCF-0546618 and CNS-0626595, and a Graduate Student Fellowship from the Notre Dame Center for Applied Mathematics. The material in this
paper was presented in part at the IEEE International Symposium on Information Theory (ISIT), Nice, France, June 2007.}
\thanks{W. Zhang is with the Communication Sciences Institute, Ming Hsieh Department of Electrical Engineering, University
of Southern California, Los Angeles, CA. Email: {\tt wenyizha@usc.edu}}
\thanks{S. Kotagiri and J. N. Laneman are with the Department of Electrical Engineering, University of Notre Dame, Notre
Dame, IN. Email: {\tt \{skotagir, jnl\}@nd.edu}}
}

\maketitle

\begin{abstract}
This paper investigates downlink transmission over a quasi-static fading Gaussian broadcast channel (BC), to model delay-sensitive applications over slowly time-varying fading channels. System performance is characterized by outage achievable rate regions. In contrast to most previous work, here the problem is studied under the key assumption that the transmitter only knows the probability distributions of the fading coefficients, but not their realizations. For scalar-input channels, two coding schemes are proposed. The first scheme is called blind dirty paper coding (B-DPC), which utilizes a robustness property of dirty paper coding to perform precoding at the transmitter. The second scheme is called statistical superposition coding (S-SC), in which each receiver adaptively performs successive decoding with the process statistically governed by the realized fading. Both B-DPC and S-SC schemes lead to the same outage achievable rate region, which always dominates that of time-sharing, irrespective of the particular fading distributions. The S-SC scheme can be extended to BCs with multiple transmit antennas.
\end{abstract}

\section{Introduction}
\label{sec:intro}

In downlink transmission, a centralized transmitter needs to simultaneously communicate with multiple receivers. Each receiver can only decode its message from its own received signal, without access to the other receivers' signals. Such systems are usually modeled as broadcast channels (BC) with Gaussian noises, which have been studied extensively since the development of superposition coding \cite{cover72:it}; see also \cite{cover98:it} and references therein for an overview of early results on BCs.

For a Gaussian BC with scalar inputs and outputs, superposition coding achieves a rate region which dominates that of time-sharing \cite{bergmans74:it-2}, and in fact yields the capacity region \cite{bergmans74:it-1}. If the transmitter and receivers are equipped with multiple antennas, the resulting vector Gaussian BC is generally non-degraded, and superposition coding turns out to be suboptimal, and dirty paper coding (DPC), originally proposed in \cite{costa83:it} for single-user Gaussian channels with Gaussian interference non-causally known at the transmitter, can be utilized to maximize the throughput \cite{caire03:it}. This observation has stimulated a series of work on vector Gaussian BCs \cite{caire03:it}-\cite{weingarten06:it}, and it has recently been shown that DPC achieves the capacity region of vector Gaussian BCs \cite{weingarten06:it}.

A central assumption in the aforementioned results is that the transmitter has perfect knowledge of the channel state information (CSI), namely, the channel gains, be they constant or random (say, due to fading). For scalar Gaussian BCs with fading, if the transmitter and all the receivers have perfect CSI, both the ergodic capacity region and the outage capacity region are known \cite{li01:it-1, li01:it-2}; however, without transmit CSI, neither is known. For ergodic fading BCs without transmit CSI, an achievable rate region has been obtained in \cite{tuninetti03:isit}.

In this paper, we investigate quasi-static fading Gaussian BCs without transmit CSI. The motivation is to model downlink transmission in delay-sensitive applications over slowly time-varying fading channels, and the lack of transmit CSI serves as the worst case for practical systems in which an adequate feedback link may not be available. Due to the non-ergodic nature of quasi-static fading, it is generally impossible for a coding scheme to achieve any strictly positive information rate under all fading realizations. We therefore focus on outage achievable rate regions, as will be formally introduced in Section \ref{sec:model}.

Lack of transmit CSI seems to pose a fundamental difficulty in broadcast settings. If the transmitter has CSI, the standard BC model is stochastically degraded conditioned upon the fading realizations, because the transmitter can sort the receivers according to their realized signal-to-noise ratios (SNR). Superposition coding is thus optimal for each channel realization, and achieves the outage capacity region when combined with dynamic power allocation \cite{li01:it-2}. However, without transmit CSI, the transmitter has no way to predict the ordering of the received signals. Conventional superposition coding therefore would not appear to be effective for this model. Generally speaking, a quasi-static fading Gaussian BC without transmit CSI belongs to the class of ``mixed channels'' \cite{han03:book}, for which no computable, single-letter characterization of the $\epsilon$-capacity region, {\it i.e.}, outage capacity region, has been obtained (cf. \cite{iwata05:ieice}).

Even though conventional superposition coding is not effective, there exist efficient approaches in terms of outage achievable rate region. In this paper, we identify two such coding schemes, and show that they both lead to the same outage achievable rate region, which always dominates that of time-sharing, irrespective of the particular fading distributions. The first scheme is called blind dirty paper coding (B-DPC), which utilizes a robustness property of DPC to perform precoding at the transmitter. The second scheme is called statistical superposition coding (S-SC), in which each receiver adaptively performs successive decoding with the process statistically governed by the realized fading.

B-DPC is a transmit-centric approach, because the transmitter needs to invoke dirty paper codes $K$ times in a progressive way, while each receiver only needs to decode its own message directly. In contrast, S-SC is the more a receive-centric approach, because the transmitter simply adds up $K$ independently coded streams as in conventional superposition coding, while each receiver (except the $K$th one) needs to execute a successive interference cancellation procedure.

The remainder of this paper is organized as follows. Section \ref{sec:model} presents the channel model and problem formulation. Section \ref{sec:main} gives the main result which characterizes the outage achievable rate region, and shows that it always dominates that of time-sharing. Sections \ref{sec:bdpc} and \ref{sec:ssc} show how the region of Section \ref{sec:main} is achieved by B-DPC and S-SC, respectively. Finally, Section \ref{sec:conclude} concludes the paper.

\section{Channel Model and Problem Formulation}
\label{sec:model}

In this section, we summarize the $K$-user scalar Gaussian BC model with quasi-static fading. The input-output relationship of the channel satisfies
\begin{eqnarray}
\label{eqn:k-user}
\rvy_k[n] = \rvh_k \rvx[n] + \rvz_k[n], \quad k = 1, \ldots, K,\;\; n = 1, \ldots, N.
\end{eqnarray}
At discrete-time index $n$, the channel takes a scalar input $\rvx[n] \in \mathbb{C}$ from the transmitter, and produces a scalar output $\rvy_k[n]
\in \mathbb{C}$ at the $k$th receiver. The channel input $\rvx[\cdot]$ has an average power constraint $P$, given as
\begin{eqnarray}
\frac{1}{N}\sum_{n = 1}^N |\rvx[n]|^2 \leq P
\end{eqnarray}
over the coding block of length $N$. The channel noise samples $\rvz_k[\cdot]$ are independent, identically distributed (i.i.d.) and circularly symmetric complex Gaussian, with mean zero and variance $N_0$, denoted $\rvz_k[\cdot] \sim \mathcal{CN}(0, N_0)$. For scalar fading channels with perfect receive CSI, as will be assumed in this paper, there is no loss of generality to consider only fading magnitudes. So we assume that the squared channel fading coefficient $\rva_k := |\rvh_k|^2$ has a probability density function (PDF) $f_k(a)$ for $a \in [0, \infty)$, and remains constant over the entire coding block so that the resulting BC is called quasi-static. We denote the cumulative distribution function (CDF) of $\rva_k$ by
$F_k(a) := \mathbb{P}[\rva_k \leq a]$, and the corresponding inverse cumulative distribution function (ICDF), or, the so-called quantile function, by $G_k(t)$. For every $t \in [0, 1]$, $G_k(t)$ is the supremum of the set $\{a: F_k(a) = t\}$.

We assume that, for each coding block, the realization of $\rvh_k$ is known perfectly at the $k$th receiver, but not at the transmitter or any other receiver. Such a situation may arise in practical systems in which receivers are able to estimate their channels with satisfactory accuracy, but the transmitter does not for lack of an adequate feedback link. Although in practice the receivers' estimate of channels is noisy due to limited channel training, we assume the receive CSI is prefect, in order to simplify analysis and provide useful insights into the more general case.

In the sequel, we will frequently make use of the average SNR defined as $\snr := P/N_0$, and without loss of generality normalize the channel equation (\ref{eqn:k-user}) such that $P = \snr$ and $N_0 = 1$.

\begin{figure}[p]
\psfrag{M}[cc]{\Large$\times$}
\psfrag{A}[cc]{\Large $+$}
\psfrag{X}{{\footnotesize $\rvx[\cdot]$}}
\psfrag{V}{\Large $\vdots$}
\psfrag{W}[cc]{$\{\rvm_1,\rvm_2,\ldots,\rvm_K\}$}
\psfrag{E}[cc]{ENC}
\psfrag{D1}[cc]{\small DEC 1}
\psfrag{D2}[cc]{\small DEC 2}
\psfrag{DK}[cc]{\small DEC K}
\psfrag{A1}[cc]{{\footnotesize$\rvh_1$}}
\psfrag{A2}[cc]{{\footnotesize $\rvh_2$}}
\psfrag{A3}[cc]{{\footnotesize $\rvh_K$}}
\psfrag{Y1}[cc]{{\footnotesize $\rvy_1[\cdot]$}}
\psfrag{Y2}[cc]{{\footnotesize $\rvy_2[\cdot]$}}
\psfrag{YK}[cc]{{\footnotesize $\rvy_K[\cdot]$}}
\psfrag{Z1}[cc]{{\footnotesize $\rvz_1[\cdot]$}}
\psfrag{Z2}[cc]{{\footnotesize $\rvz_2[\cdot]$}}
\psfrag{ZK}[cc]{{\footnotesize $\rvz_K[\cdot]$}}
\psfrag{M1}[cc]{{\footnotesize $\hat{\rvm}_1$}}
\psfrag{M2}[cc]{{\footnotesize $\hat{\rvm}_2$}}
\psfrag{M3}[cc]{{\footnotesize $\hat{\rvm}_K$}}
\epsfxsize=5in
\epsfclipon
\centerline{\epsffile{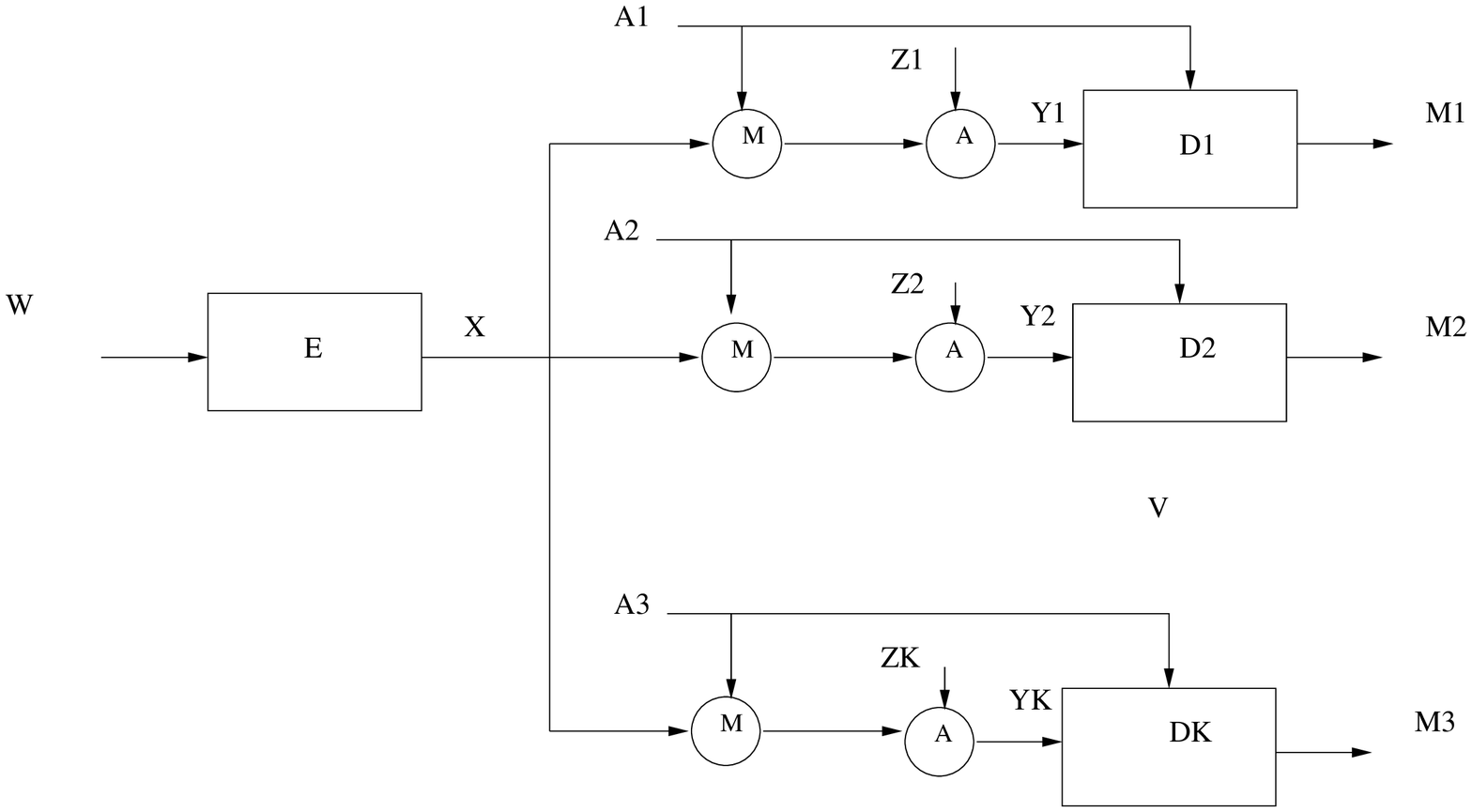}}
\caption{Block diagram for a $K$-user scalar Gaussian broadcast channel with quasi-static fading.}
\label{fig:illustration}
\end{figure}

For one coding block, the encoder maps $K$ mutually independent messages, each for one individual user, altogether into a codeword of length $N$, {\it i.e.},
\begin{eqnarray}
\{\rvx[n]\}_{n = 1}^N = \varphi^{(N)}\left(\left\{\rvm_k\right\}_{k = 1}^K\right).
\end{eqnarray}
Note that the encoding function $\varphi^{(N)}(\cdot)$ does not depend upon the realization of the fading coefficients $\{\rva_k\}_{k = 1}^K$.
The $k$th message, $\rvm_k$, is uniformly chosen from $\{1, \ldots, \left\lceil \exp(N R_k)\right\rceil\}$
where $R_k \geq 0$ is the target rate for the $k$th user. The $k$th decoder maps its received signal along with its fading coefficient into a message index in $\{1, \ldots, \left\lceil \exp(N R_k)\right\rceil\}$, as
\begin{eqnarray}
\hat{\rvm}_k = \psi^{(N)}_k\left(\left\{\rvy_k[n]\right\}_{n = 1}^N, \rva_k\right).
\end{eqnarray}

For a sequence of encoder-decoders tuples $\{\varphi^{(N)}(\cdot), \psi^{(N)}_1(\cdot), \ldots, \psi^{(N)}_K(\cdot)\}$,
indexed by the coding block length $N$, and an outage probability vector
$\underline{\epsilon} = (\epsilon_1, \ldots, \epsilon_K) \in [0, 1]^K$,
we say that a rate vector $\underline{R} = (R_1, \ldots, R_K)$ is $\underline{\epsilon}$-outage
achievable if the outage probability for the $k$th user
\begin{eqnarray*}
\limsup_{N \rightarrow \infty} \mathbb{P}\left[\hat{\rvm}_k = \psi^{(N)}_k\left(\left\{\rvy_k[n]\right\}_{n = 1}^N, \rva_k\right) \neq \rvm_k\right] \leq \epsilon_k,
\end{eqnarray*}
simultaneously for $k = 1, \ldots, K$. The $\underline{\epsilon}$-outage capacity region $\mathcal{C}(\snr, \underline{\epsilon})$
is then defined as the closure of the set of all the $\underline{\epsilon}$-outage achievable rate vectors for all possible encoder-decoders tuples, under the input power constraint (cf. \cite{li01:it-2}).

\section{An Outage Achievable Rate Region}
\label{sec:main}

For the channel model introduced in Section \ref{sec:model}, we have the following result.
\begin{prop}
\label{prop:k-user}
For the $K$-user quasi-static fading scalar Gaussian BC without transmit CSI, and a given outage probability vector $\underline{\epsilon}$,
sorting the indexes of the $K$ receivers such that
$G_1(\epsilon_1) \geq G_2(\epsilon_2) \geq \ldots \geq G_K(\epsilon_K)$,
an $\underline{\epsilon}$-outage achievable rate region is given by
\begin{eqnarray}
\label{eqn:R-inner}
\mathcal{R}^\ast(\snr, \underline{\epsilon}) := \left\{\underline{R}: \exists \underline{\gamma} = (\gamma_1, \ldots, \gamma_K) \in [0, 1]^K, \sum_{k = 1}^K \gamma_k = 1, \mbox{s.t.}\; R_k < R^\ast_k(\snr, \underline{\gamma}, \epsilon_k), \forall k = 1, \ldots, K\right\},
\end{eqnarray}
where
\begin{eqnarray}
\label{eqn:Rk}
R^\ast_k(\snr, \underline{\gamma}, \epsilon_k) &:=& \log\left(1 + 
\frac{G_k({\epsilon}_k) \gamma_k\snr}
{G_k({\epsilon}_k)\cdot (\sum_{i = 1}^{k - 1}\gamma_i)\snr + 1}
\right).
\end{eqnarray}
\end{prop}
{\it Proof}: We provide two different proofs of the achievability of $\mathcal{R}(\snr, \underline{\epsilon})$ in Sections \ref{sec:bdpc} and \ref{sec:ssc}, respectively. {\bf Q.E.D.}

We emphasize that, in Proposition \ref{prop:k-user}, the $K$ users are sorted based upon the values of $G_k(\epsilon_k)$, $k = 1, \ldots, K$. This is a crucial condition. As will be demonstrated in Section \ref{sec:bdpc}, for any arbitrary ordering of the $K$ users, we can obtain an $\underline{\epsilon}$-outage achievable rate region given by (\ref{eqn:R-inner}). However, the resulting region is largest only for the particular ordering specified here.

\subsection*{Comparison with Time-Sharing}

If we employ time-sharing to decompose a BC into $K$ non-interfering, single-user channels with time-sharing vector $\underline{\mu} = (\mu_1, \ldots, \mu_K) \in [0, 1]^K, \sum_{k = 1}^K \mu_k = 1$, and further allow power allocation among these $K$ channels with power allocation vector $\underline{\eta} = (\eta_1, \ldots, \eta_K) \in [0, \infty)^K$ such that $\sum_{k = 1}^K \mu_k \eta_k = 1$, then it follows that we can achieve an $\underline{\epsilon}$-outage achievable rate region given by
\begin{eqnarray}
\mathcal{R}^\mathrm{td}(\snr, \underline{\epsilon}) := \left\{
\underline{R}: \exists \underline{\mu}, \underline{\eta}, \mbox{s.t.}\; R_k < R^\mathrm{td}_k(\snr, \mu_k, \eta_k, \epsilon_k), \forall k = 1, \ldots, K
\right\},
\end{eqnarray}
where
\begin{eqnarray}
R^\mathrm{td}_k(\snr, \mu_k, \eta_k, \epsilon_k) &:=& \mu_k
\cdot\log\left(1 + G_k({\epsilon}_k) \eta_k\snr\right).
\end{eqnarray}

In order to compare $\mathcal{R}^\ast(\snr, \underline{\epsilon})$ and $\mathcal{R}^\mathrm{td}(\snr, \underline{\epsilon})$, it is useful to introduce the following memoryless Gaussian BC without fading,
\begin{eqnarray}
\label{eqn:bc-nonfading}
\tilde{\rvy}_k[i] = \sqrt{G_k({\epsilon}_k)} \tilde{\rvx}[i] +
\tilde{\rvz}_k[i],\quad k = 1, \ldots, K,\;\; i = 1, \ldots, n,
\end{eqnarray}
with $\tilde{\rvz}_k[\cdot] \sim \mathcal{CN}(0, 1)$, and with the same average power constraint $\rho$ as in the original quasi-static fading BC (\ref{eqn:k-user}).
We then notice that $\mathcal{R}^\ast(\snr, \underline{\epsilon})$ coincides with the capacity region of this Gaussian BC (\ref{eqn:bc-nonfading}), while $\mathcal{R}^\mathrm{td}(\snr, \underline{\epsilon})$ corresponds to its rate region achieved by time-sharing. Therefore we conclude that
$\mathcal{R}^\ast(\snr, \underline{\epsilon}) \supseteq \mathcal{R}^\mathrm{td}(\snr, \underline{\epsilon})$, and note that the two regions coincide if and only if $G_1(\epsilon_1) = G_2(\epsilon_2) = \ldots = G_K(\epsilon_K)$ (cf. \cite{bergmans74:it-2}). That is, Proposition \ref{prop:k-user} yields an $\underline{\epsilon}$-outage achievable rate region that always contains that of time-sharing.

\begin{figure}[p]
\psfrag{xlabel}{{$R_1$} (nats)}
\psfrag{ylabel}{{$R_2$} (nats)}
\psfrag{prop1}{{$\mathcal{R}^\ast(\snr, \underline{\epsilon})$}}
\psfrag{tdpa}{{$\mathcal{R}^\mathrm{td}(\snr, \underline{\epsilon})$}}
\epsfxsize=5.0in
\epsfclipon
\centerline{\epsffile{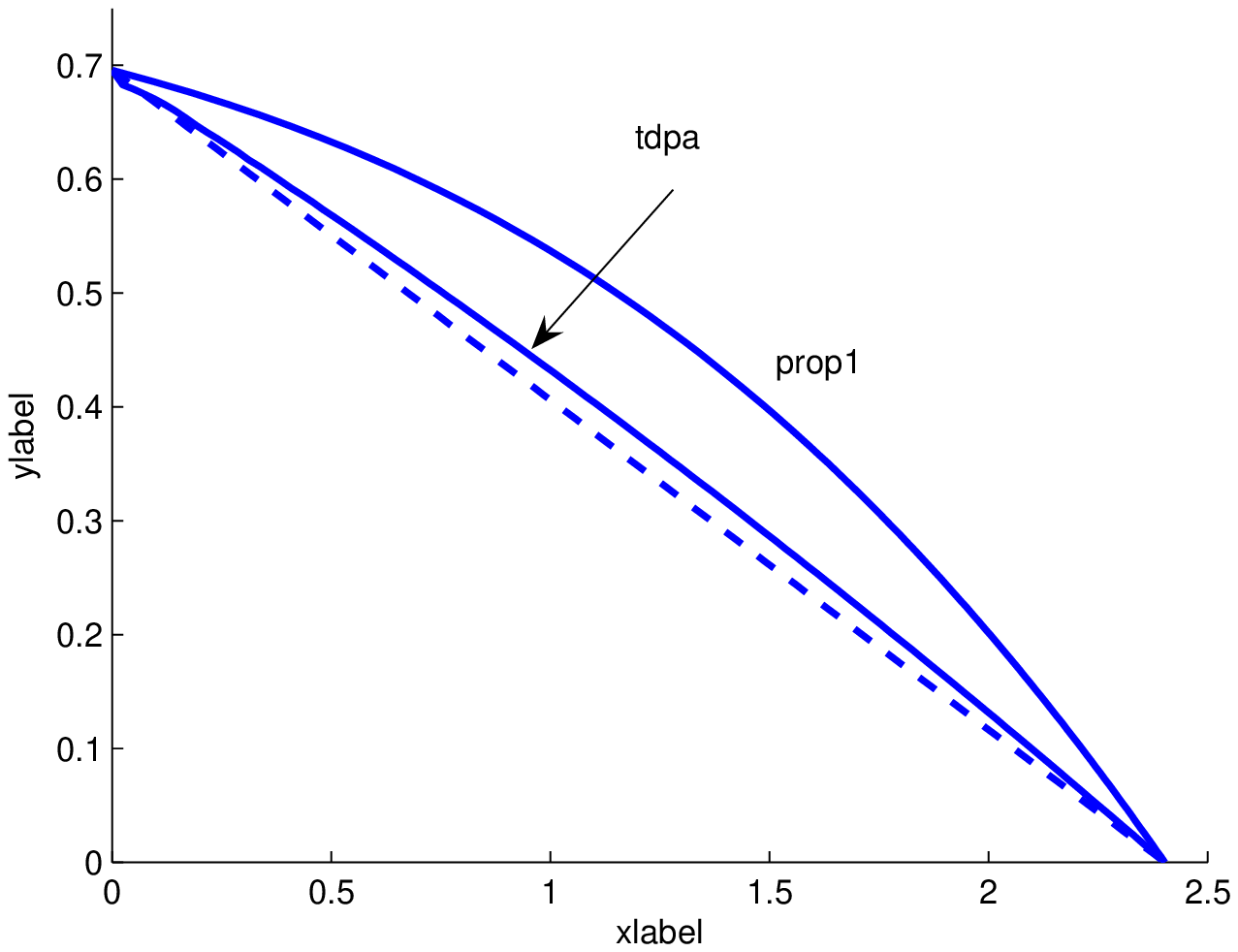}}
\caption{Comparison of outage achievable rate regions given in Proposition \ref{prop:k-user} and achieved by time-sharing with power allocation (solid) and without power allocation (dashed).}
\label{fig:good_dpc}
\end{figure}

For illustration, let us examine an example with two receivers. Both receivers experience Rayleigh fading, {\it i.e.}, $\rva_1, \rva_2$ are exponential random variables. We assume that the two receivers are under a near-far situation, with $\mathbf{E}[\rva_1] = 10$ and $\mathbf{E}[\rva_2] = 1$. The target outage probability vector is $\underline{\epsilon} = [0.01 \;\; 0.01]$, and the average power constraint $\rho$ is $20$dB. From these parameters, we find that $G_1(\epsilon_1) = 10\times\log(1/0.99) \approx 0.1$ and $G_2(\epsilon_2) = \log(1/0.99) \approx 0.01$, respectively. Figure \ref{fig:good_dpc} depicts the $\underline{\epsilon}$-outage achievable rate regions
$\mathcal{R}^\ast(\snr, \underline{\epsilon})$ and $\mathcal{R}^\mathrm{td}(\snr, \underline{\epsilon})$, from which it is clear that $\mathcal{R}^\ast(\snr, \underline{\epsilon})$ contains $\mathcal{R}^\mathrm{td}(\snr, \underline{\epsilon})$.

\section{Blind Dirty Paper Coding (B-DPC)}
\label{sec:bdpc}

In this section, we present the first coding scheme that achieves $\mathcal{R}^\ast(\snr, \underline{\epsilon})$ in Proposition \ref{prop:k-user}. We first introduce a variant of the ``writing on dirty paper'' (WDP) problem and observe a robustness property of B-DPC, then utilize this property to establish the achievability of $\mathcal{R}^\ast(\snr, \underline{\epsilon})$.

\subsection{Blind DPC and a Robustness Property}
\label{subsec:robustness}

\begin{figure}[p]
\psfrag{X}{$\rvx[\cdot]$}
\psfrag{S1}{$\rvs_1[\cdot]$}
\psfrag{S2}{$\rvs_2[\cdot]$}
\psfrag{Z}{$\rvz[\cdot]$}
\psfrag{sqrtA}{$\sqrt{\rva}$}
\psfrag{Y}{$\rvy[\cdot]$}
\psfrag{Mess}{$\rvm$}
\psfrag{hatMess}{$\hat{\rvm}$}
\epsfxsize=5in
\epsfclipon
\centerline{\epsffile{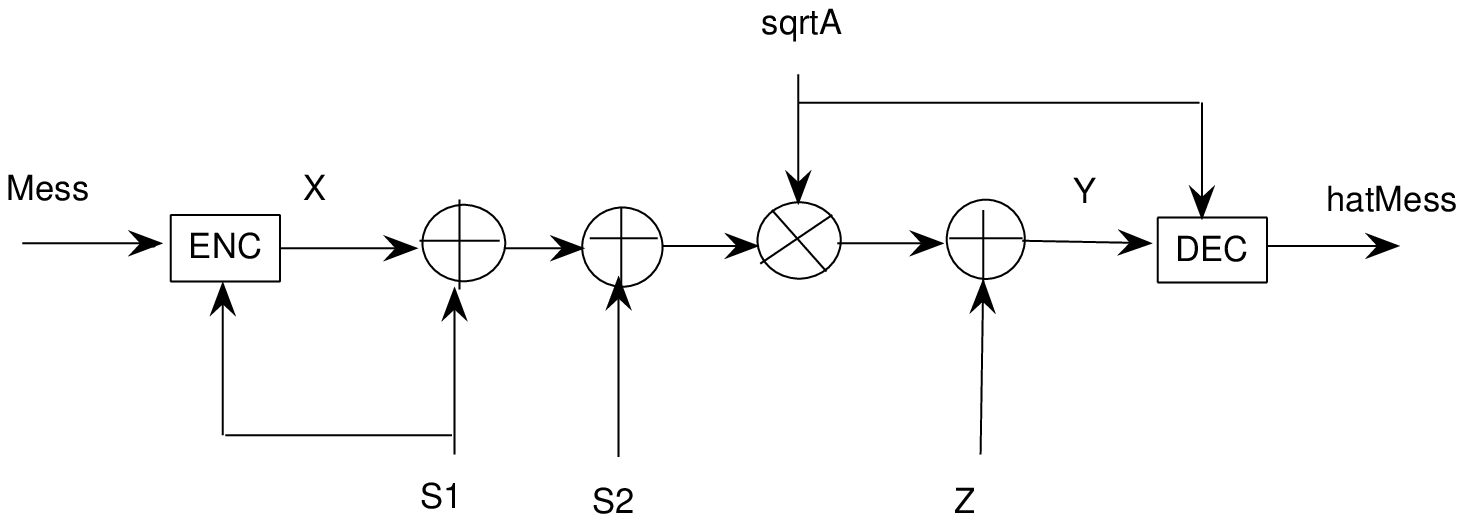}}
\caption{The variant of the ``writing on dirty paper'' problem considered in Section \ref{subsec:robustness}.}
\label{fig:wdp}
\end{figure}

Consider a variant of the WDP problem illustrated in Figure \ref{fig:wdp}. The channel law satisfies
\begin{eqnarray}
\label{eqn:wdp}
\rvy[n] = \sqrt{\rva} \cdot (\rvx[n] + \rvs_1[n] + \rvs_2[n]) + \rvz[n],~~ n = 1, \ldots, N,
\end{eqnarray}
with i.i.d. additive noise $\rvz[\cdot] \sim \mathcal{CN}(0, N_0)$, and i.i.d. interference signals $\rvs_1[\cdot] \sim \mathcal{CN}(0, Q_1)$ and $\rvs_2[\cdot] \sim \mathcal{CN}(0, Q_2)$. The input $\rvx[\cdot]$ has an average power constraint $P$. The transmitter has full access to $\rvs_1$ non-causally, but neither the transmitter nor the receiver has access to $\rvs_2$; thus $\rvs_2$ acts as a (faded) noise. The fading, or resizing, random variable $\rva$ has a PDF $f(a)$ for $a \in [0, \infty)$, and remains constant over the entire coding block. Furthermore, $\rva$ is known at the receiver but not at the transmitter.

We note that, (\ref{eqn:wdp}) reduces to the original WDP problem if and only if $\rva$ is a constant with probability one. For general distributions on $\rva$, the channel SNR $\rva P/(\rva Q_2 + N_0)$ is a random variable unknown to the transmitter due to its lack of knowledge of $\rva$. Therefore it is impossible for the transmitter to dynamically adapt its DPC scheme according to the channel realization. Nevertheless, we can still apply DPC, with a linear precoding coefficient $\alpha$ chosen independent of $\rva$, to generate the auxiliary random variable $\rvU = \rvx + \alpha \rvs_1$. We call this approach ``blind'' dirty paper coding (B-DPC).

Following the DPC encoding and decoding procedures in \cite{costa83:it}, and noting that the channel fading only affects the noise variance at the decoder, we can find that the achievable rate conditioned on $\rva$ is the random variable
\begin{eqnarray}
\label{eqn:cond-rate}
\rvJ(\alpha, \rva) :=
\log \frac{P[\rva (P + Q_1 + Q_2) + N_0]}{(1 - \alpha)^2 \rva PQ_1 + (P + \alpha^2 Q_1)(\rva Q_2 + N_0)}.
\end{eqnarray}
For every target rate $R \geq 0$, (\ref{eqn:cond-rate}) thus enables us to evaluate the outage probability $\mathbb{P}\left[\rvJ(\alpha, \rva) \leq R\right]$, {\it i.e.}, the probability that the realization of $\rva$ makes the achievable rate $\rvJ(\alpha, \rva)$ insufficient to support the target rate $R$. We further adjust the linear precoding coefficient $\alpha$ to minimize the outage probability. After manipulations, we find that the minimizer of $\mathbb{P}\left[\rvJ(\alpha, \rva) \leq R\right]$ is
\begin{eqnarray}
\label{eqn:alpha-opt}
\alpha^\ast = 1 - e^{-R},
\end{eqnarray}
and that the corresponding minimum outage probability is
\begin{eqnarray}
\label{eqn:min-outage}
\min_{\alpha} \mathbb{P}\left[\rvJ(\alpha, \rva) \leq R\right] = \mathbb{P}\left[
R \geq \log \left(1 + \frac{P \rva}{Q_2 \rva + N_0}\right)
\right].
\end{eqnarray}

From (\ref{eqn:min-outage}), we observe that the minimum outage probability of B-DPC coincides with the minimum outage probability if the receiver also knows $\rvs_1[\cdot]$ and thus can eliminate $\sqrt{\rva} \rvs_1[\cdot]$ from the received signal. Therefore B-DPC is outage-optimal, regardless of the specific distribution of $\rva$. It is also interesting to note that the optimal choice of $\alpha$ depends upon the target rate $R$. We may introduce a virtual channel SNR $\snr^\ast$ satisfying $R = \log(1 + \rho^\ast)$, and rewrite (\ref{eqn:alpha-opt}) as $\alpha^\ast = {\snr^\ast}/{(1 + \snr^\ast)}$. So for a given target rate $R$, the optimal strategy for the transmitter is to treat the channel as if it is realized to just be able to support this rate.

The optimality of B-DPC can be explained by a coincidence argument as follows. The conditional achievable rate (\ref{eqn:cond-rate}) is a function of two variables, $\alpha$ and $\rva$, and is monotonically increasing with $\rva$ for every $\alpha$. On the other hand, for $\rva$ known to the transmitter, the choice of $\alpha$ maximizing $\rvJ(\alpha, \rva)$ is given by $\alpha^{\mathrm{DPC}}(\rva) := \rva P/(\rva P + \rva Q_2 + N_0)$. Therefore, for a given target rate $R$, if we solve the equation $\rvJ(\alpha^{\mathrm{DPC}}(\rva), \rva) = R$ which has the unique solution $\rva = a^\ast$, and choose $\alpha^\ast = \alpha^\mathrm{DPC}(a^\ast)$ in B-DPC, we can guarantee that for every fading realization $\rva < a^\ast$, the target rate $R$ is always achievable.

\subsection{Proof of Proposition \ref{prop:k-user} via B-DPC}
\label{subsec:proof}

We now proceed to proving Proposition \ref{prop:k-user} using B-DPC. For every fixed $\underline{\gamma}$, we need to show that all rate vectors $\underline{R}$ satisfying (\ref{eqn:R-inner}) are achievable. Consider the $k$th receiver, and rewrite its channel as
\begin{eqnarray}
\label{eqn:bdpc-kchannel}
\rvy_k[n] = \sqrt{\rva_k} \rvx_k[n] + \sqrt{\rva_k} \sum_{l > k} \rvx_l[n] +
(\sqrt{\rva_k} \sum_{m < k} \rvx_m[n] + Z_k[n]),\quad n = 1, \ldots, N.
\end{eqnarray}
In (\ref{eqn:bdpc-kchannel}), the encoder function $\varphi^{(N)}\left(\left\{\rvm_k\right\}_{k = 1}^K\right)$ is additive such that
\begin{eqnarray}
\left\{\rvx[n]\right\}_{n = 1}^N = \sum_{k = 1}^K \varphi^{(N)}_k\left(\left\{\rvm_i\right\}_{i = k}^K\right),
\end{eqnarray}
and we denote
\begin{eqnarray}
\left\{\rvx_k[n]\right\}_{n = 1}^N = \varphi^{(N)}_k\left(\left\{\rvm_i\right\}_{i = k}^K\right), \quad k = 1, \ldots, K.
\end{eqnarray}
We encode $\rvm_k$ into $\left\{\rvx_k[n]\right\}_{n = 1}^N$ following B-DPC with average power $\gamma_k\snr$, by treating $\sum_{l > k} \rvx_l[\cdot]$ as the non-causally known interference, and by treating $(\sqrt{\rva_k} \sum_{m < k} \rvx_m[\cdot] + \rvz_k[\cdot])$ as noise. The encoded signal $\{\rvx_k[n]\}_{n = 1}^N$ thus contains i.i.d. $\mathcal{CN}(0, \gamma_k \snr)$ components, which are further mutually independent with any other $\rvx_{k^\prime}[\cdot]$, $\forall k^\prime \neq k$. From the discussion in Section \ref{subsec:robustness}, if we choose the linear precoding coefficient in B-DPC as $\alpha^\ast_k = 1 - e^{-R_k}$ for a target rate $R_k$, the resulting outage probability of the $k$th receiver is
\begin{eqnarray}
\label{eqn:outage-proof}
\mathbb{P}^{(\mathrm{out})}_k := \mathbb{P}\left[
\rva_k \leq \frac{e^{R_k} - 1}{\gamma_k \snr - (e^{R_k} - 1) \sum_{m < k}\gamma_m \snr}
\right].
\end{eqnarray}

Alternatively, for a given target outage probability $\epsilon_k$ for the $k$th receiver, it follows from (\ref{eqn:outage-proof}) that the maximum achievable rate $R_k$ should satisfy
\begin{eqnarray*}
\frac{e^{R_k} - 1}{\gamma_k \snr - (e^{R_k} - 1) \sum_{m < k}\gamma_m\snr} < G_k(\epsilon_k),
\end{eqnarray*}
which gives rise to
\begin{eqnarray}
R_k < \log \left(
1 + \frac{G_k(\epsilon_k) \gamma_k\snr}{G_k(\epsilon_k) \sum_{m < k}\gamma_m \snr + 1}
\right),
\end{eqnarray}
corresponding to (\ref{eqn:Rk}) for the fixed $\underline{\gamma}$. As we exhaust all the possible $\underline{\gamma}$, we obtain the rate region $\mathcal{R}^\ast(\snr, \underline{\epsilon})$ as given by (\ref{eqn:R-inner}). This concludes the proof of Proposition \ref{prop:k-user}.

\subsection{Extension to Receivers with Multiple Antennas}
\label{subsec:simo}

Proposition \ref{prop:k-user} readily extends to the case in which each receiver has multiple antennas. This stems from the fact that DPC \cite{costa83:it}
can be extended (by directly applying the general results in \cite{gelfand80:pcit}) to single-input, multiple-output (SIMO) Gaussian channels. Analogously, B-DPC still attains robustness without transmit CSI, and the steps in Sections \ref{subsec:robustness} and \ref{subsec:proof} carry through.

Consider a $K$-user quasi-static fading scalar-input Gaussian BC without transmit CSI, with the $k$th receiver equipped with $m_k$ receive antennas receiving
\begin{eqnarray}
\label{eqn:simo}
\rvy_{k, m}[n] = \rvh_{k, m} \rvx[n] + \rvz_{k, m}[n], \quad m = 1, \ldots, m_k,\;\; n = 1, \ldots, N.
\end{eqnarray}
The i.i.d. additive noise vector $\bigl[\rvz_{k, 1}[\cdot], \ldots, \rvz_{k, m_k}[\cdot]\bigr]^\mathrm{T} \sim \mathcal{CN}(\mathbf{0}, \mathbf{I}_{m_k \times m_k})$. The input $\rvx[\cdot]$ satisfies average power constraint $\snr$. The complex-valued random variable $\rvh_{k, m}$ denotes the fading coefficient for
the $m$th receive antenna of the $k$th receiver. Here note that for vector fading channels, we need to take into consideration the complex-valued fading coefficients. We have the following result.

\begin{cor}
\label{cor:simo}
For the channel model (\ref{eqn:simo}), B-DPC achieves an $\underline{\epsilon}$-outage achievable rate region identical to that described by $\mathcal{R}^\ast(\snr, \underline{\epsilon})$ for the $K$-user quasi-static fading scalar Gaussian BC model (\ref{eqn:k-user}), with $\rva_k$ replaced by $\sum_{m = 1}^{m_k} |\rvh_{k, m}|^2$.
\end{cor}

{\it Case Study: Receivers with Two Antennas of Spatially Correlated Rayleigh Fading}

\begin{figure}[p]
\psfrag{xlabel}{$R_1$ (nats)}
\psfrag{ylabel}{$R_2$ (nats)}
\epsfxsize=5in
\epsfclipon
\centerline{\epsffile{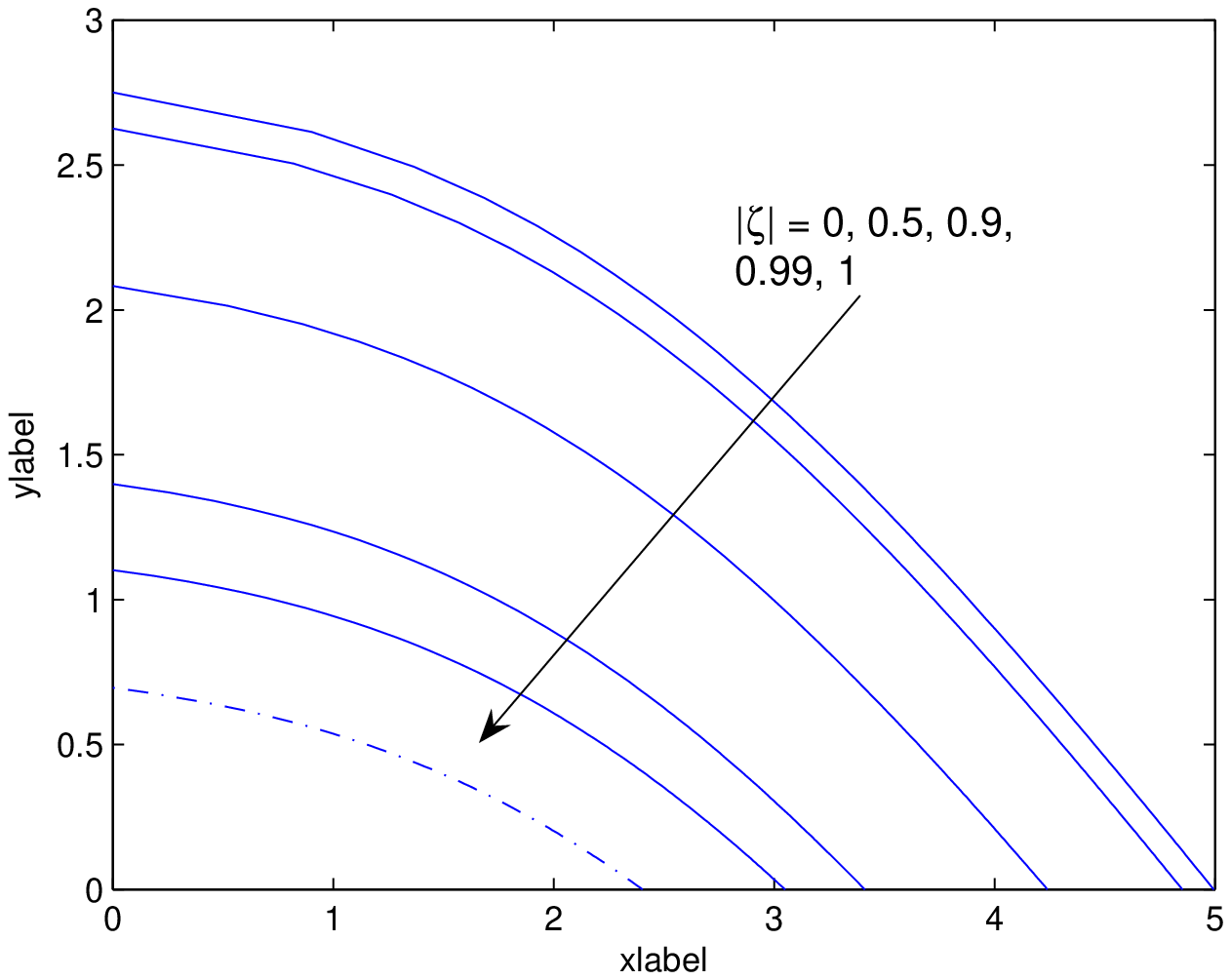}}
\caption{Outage achievable rate regions for a two-user BC with each receiver equipped with two antennas experiencing spatially correlated Rayleigh fading. For comparison, in dashed-dot curve we plot the outage achievable rate region with each receiver equipped with a single receive antenna.}
\label{fig:simo}
\end{figure}

In practical downlink systems, the physical size of receivers is usually limited. Consequently, the number of receive antennas is typically small and spatial correlation exists among them. Here we examine the case of two receivers, each equipped with two antennas experiencing Rayleigh fading. For each receiver, the fading coefficients of the two receive antennas are correlated with correlation coefficient $\zeta \in [-1, 1]$. We assume that the two receivers are under a near-far situation, with the mean of each fading coefficient of the first receiver being $10$ and that of the second being $1$. The target outage probability vector is $\underline{\epsilon} = [0.01 \; 0.01]$, and the average power constraint $\rho$ is $20$dB. Figure \ref{fig:simo} depicts the $\underline{\epsilon}$-outage achievable rate regions $\mathcal{R}^\ast(\rho, \underline{\epsilon})$, for different values of the spatial correlation coefficient $\zeta$. It is clearly illustrated that multiple receive antennas, even moderately correlated, substantially enlarge the outage achievable rate region.

\section{Statistical Superposition Coding (S-SC)}
\label{sec:ssc}

As we know, and the robustness property of B-DPC exemplifies, outage probability relates more to the fading statistics rather to individual realizations. We therefore are motivated to revisit superposition coding, focusing on its statistical properties in the context of quasi-static fading. As will be shown in this section, a modified superposition coding scheme, called statistical superposition coding (S-SC), also achieves the $\underline{\epsilon}$-outage achievable rate region $\mathcal{R}^\ast(\snr, \underline{\epsilon})$ given by Proposition \ref{prop:k-user}.

\subsection{Encoding and Decoding Procedures for S-SC}

\noindent \textbf{Encoding:}
The encoding part of S-SC is identical to conventional superposition coding for a scalar Gaussian BC \cite{cover72:it}. Fix a power allocation vector $\underline{\gamma} = (\gamma_1, \ldots, \gamma_K) \in [0, 1]^K$ satisfying $\sum_{k = 1}^K \gamma_k = 1$. The channel inputs are again generated as $\rvx[\cdot] = \sum_{k = 1}^K \rvx_k[\cdot]$, where the i.i.d. $\rvx_k[\cdot] \sim \mathcal{CN}(0, \gamma_k\snr)$ encodes the message $\rvm_k$ for the $k$th receiver. We note that, however, the $K$ signal components $\{\rvx_k[\cdot]\}_{k = 1}^K$ are generated independently, with no dependence as in B-DPC.

\noindent \textbf{Decoding:}
Consider the decoding procedure at the $k$th receiver, with its channel written as
\begin{eqnarray}
\rvy_k[n] = \sqrt{\rva_k} \sum_{k = 1}^K \rvx_k[n] + \rvz_k[n],\quad n = 1, \ldots, N.
\end{eqnarray}
\begin{itemize}
\item In the first step, the decoder attempts to decode $\rvm_K$, the message for the $K$th receiver, by treating $\sqrt{\rva_k} \sum_{l = 1}^{K - 1}\rvx_l[\cdot]$ as noise. Due to the quasi-static nature of the channel, the decoder may either successfully decode $\rvm_K$, and thus reliably reconstruct $\{\rvx_K[n]\}_{n = 1}^N$, or experience an outage at this stage.

\item The second decoding step has two possibilities. If $\rvm_K$ has been decoded successfully, the decoder subtracts $\sqrt{\rva_k} \rvx_K[\cdot]$ from $\rvy_k[\cdot]$, and proceeds to decode $\rvm_{K - 1}$ by treating $\sqrt{\rva_k} \sum_{l = 1}^{K - 2}\rvx_l[\cdot]$ as noise; otherwise, the decoder attempts to decode $\rvm_{K - 1}$ by treating $\sqrt{\rva_k} \sum_{l = 1}^{K - 2}\rvx_l[\cdot]$ together with $\sqrt{\rva_k} \rvx_K[\cdot]$ as noise.

\item Continuing the step-wise decoding procedure, when the decoder at the $k$th receiver turns to decode its own message $\rvm_k$, it has already successfully decoded the messages for a random subset of the other receivers with indexes larger than $k$. The decoder thus subtracts from $\rvy_k[\cdot]$ the signals for these other receivers, and decodes $\rvm_k$ by treating all the remaining undecoded signals as noise.
\end{itemize}

We note that, in the described decoding procedure, the decoder can only cancel the interfering signals of a random subset of receivers, rather than those of all
the ``more degraded'' receivers as in conventional superposition coding. This is why we call the scheme statistical superposition coding.

\subsection{Proof of Proposition \ref{prop:k-user} via S-SC}

In the proof, it suffices to show that for any fixed power allocation vector $\underline{\gamma}$, the $k$th receiver employing S-SC achieves an outage probability no larger than $\epsilon_k$, $k = 1, \ldots, K$, if the target rate vector $\underline{R}$ satisfies
\begin{eqnarray}
\label{eqn:rate-condition-ssc}
R_k < \log \left(
1 + \frac{G_k(\epsilon_k) \gamma_k\snr}{G_k(\epsilon_k) \sum_{m < k}\gamma_m \snr + 1}
\right).
\end{eqnarray}
We prove this statement by induction.

First, the statement obviously holds true for the $K$th receiver.

Next, assuming that the statement holds true for all receivers with indexes larger than $k$, consider the $k$th receiver with $k \leq K - 1$. Let us introduce a decoding-indicator for the $k$th receiver, which is a length-$(K - k + 1)$ random vector $\underline{\rvd}^{(k)} \in \{0, 1\}^{K - k + 1}$, with $l$th element $\rvd^{(k)}_l = 1$ if the decoder at the $k$th receiver has successfully decoded $\rvm_{K + 1 - l}$, and $\rvd^{(k)}_l = 0$ otherwise. For example, consider a three-user BC with the first receiver ($k = 1$) obtaining $\underline{\rvd}^{(1)} = [1 \;\;0\;\; 1]$ in one particular channel realization. This means that the first receiver has first successfully decoded the message $\rvm_3$, then experienced an outage in attempting to decode $\rvm_2$, and finally decoded its own message $\rvm_1$ successfully. In general, any $(0, 1)$-vector of appropriate length can be realized as a valid decoding-indicator due to the randomness of the fading; however, the situation is considerably simplified under the condition in Proposition \ref{prop:k-user}, namely, the indexes of the $K$ receivers are sorted such that $G_1(\epsilon_1) \geq G_2(\epsilon_2) \geq \ldots \geq G_K(\epsilon_K)$.

Under the condition of Proposition \ref{prop:k-user}, we claim that, if $\rvd^{(k)}_l = 1$ for some $l$, then $\rvd^{(k)}_{l^\prime} = 1$ for all $l^\prime \geq l + 1$. In words, if the $k$th receiver successfully decodes the message for the $l$th ($l \geq k$) receiver, then it must have successfully decoded the messages for all the receivers with indexes larger than $l$. For example, the decoding-indicator $\underline{\rvd}^{(1)} = [1 \;\;0\;\; 1]$ is impossible in this case, but $\underline{\rvd}^{(1)} = [1 \;\;1 \;\; 1], [1 \;\;1\;\; 0], [1 \;\;0\;\; 0]$, or $[0 \; 0 \; 0]$ are possible decoding-indicators.

We prove the claim by contradiction. Let us assume that there exists an execution of S-SC at the $k$th receiver with $\underline{\rvd}^{(k)}$, in which $\rvd^{(k)}_{l^\prime} = 0$ is the first zero element scanning from left to right, and $\rvd^{(k)}_l = 1$ for some $l > l^\prime$ is located to the right of $\rvd^{(k)}_{l^\prime}$ in $\underline{\rvd}^{(k)}$. Since $\rvd^{(k)}_{l^\prime} = 0$ is the first zero element in $\underline{\rvd}^{(k)}$, all the messages for the receivers with index larger than $(K + 1 - l^\prime)$ have been successfully decoded and thus eliminated from the received signal, before decoding $\rvm_{K + 1 - l^\prime}$. We therefore have
\begin{eqnarray}
\label{eqn:ssc-proof-1}
\log\left(1 + \frac{\rva_k \gamma_{K + 1 - l^\prime}\snr}
{1 + \rva_k \sum_{j = 1}^{K - l^\prime}\gamma_j \snr}\right)
 \leq R_{K + 1 - l^\prime}.
\end{eqnarray}
Meanwhile, since our induction assumes that the target rate of the $(K + 1 - l^\prime)$-th receiver satisfies (\ref{eqn:rate-condition-ssc}), {\it i.e.},
\begin{eqnarray}
\label{eqn:ssc-proof-2}
R_{K + 1 - l^\prime} < \log\left(1 + \frac{G_{K + 1 - l^\prime}(\epsilon_{K + 1 - l^\prime}) \gamma_{K + 1 - l^\prime} \snr}{1 + G_{K + 1 - l^\prime}(\epsilon_{K + 1 - l^\prime}) \sum_{j = 1}^{K - l^\prime}\gamma_j\snr}\right).
\end{eqnarray}
Comparing (\ref{eqn:ssc-proof-1}) and (\ref{eqn:ssc-proof-2}), we find that the channel fading realization $\rva_k$ must satisfy $\rva_k < G_{K + 1 - l^\prime}(\epsilon_{K + 1 - l^\prime})$.

On the other hand, $\rvd^{(k)}_l = 1$ implies
\begin{eqnarray}
\log\left(1 + \frac{\rva_k \gamma_{K + 1 - l}\snr}{1 +
\rva_k \sum_{j = 1}^{K - l}\gamma_j \snr + \rva_k \sum_{j = 1}^l
\gamma_{K + 1 - j} (1 - \rvd^{(k)}_j) \snr}\right) > R_{K + 1 - l},
\end{eqnarray}
where $\rva_k \sum_{j = 1}^l \gamma_{K + 1 - j} (1 - \rvd^{(k)}_j) \snr \geq 0$ accounts for the effect of those undecoded messages subject to outage in the previous S-SC decoding steps. So we further get
\begin{eqnarray}
\label{eqn:ssc-proof-3}
\log\left(1 + \frac{\rva_k \gamma_{K + 1 - l}\snr}{1 +
\rva_k \sum_{j = 1}^{K - l}\gamma_j \snr}\right) > R_{K + 1 - l}.
\end{eqnarray}
Meanwhile, since the induction should hold true for any rate vector satisfying (\ref{eqn:rate-condition-ssc}), we can choose an arbitrarily small $\delta > 0$ such that
\begin{eqnarray}
\label{eqn:ssc-proof-4}
R_{K + 1 - l} > \log\left(1 + \frac{G_{K + 1 - l}(\epsilon_{K + 1 - l}) \gamma_{K + 1 - l} \snr}{G_{K + 1 - l}\sum_{j = 1}^{K - l} \gamma_j \snr + 1}\right) - \delta.
\end{eqnarray}
Comparing (\ref{eqn:ssc-proof-3}) and (\ref{eqn:ssc-proof-4}), we find that $\rva_k$ must satisfy $\rva_k \geq G_{K + 1 - l}(\epsilon_{K + 1 - l})$. Combining the two bounds on $\rva_k$, we obtain
\begin{eqnarray*}
G_{K + 1 - l}(\epsilon_{K + 1 - l}) \leq \rva_k < G_{K + 1 - l^\prime}(\epsilon_{K + 1 - l^\prime}),
\end{eqnarray*}
which is in contradiction with the condition $G_1(\epsilon_1) \geq G_2(\epsilon_2) \geq \ldots \geq G_K(\epsilon_K)$. So the claim is proved.

Having established the claim regarding the structure of decoding-indicators, we are ready to complete the proof of Proposition \ref{prop:k-user}, by evaluating the probability that the $k$th receiver does not experience an outage in decoding its own message $\rvm_k$. From our claim, the occurrence of this event implies that
the messages $\{\rvm_K, \rvm_{K - 1}, \ldots, \rvm_{k + 1}\}$ all have been successfully decoded. It is then follows that for every $R_k$ satisfying
(\ref{eqn:rate-condition-ssc}), the outage probability for decoding $\rvm_k$ is no larger than $\epsilon_k$. By induction, this concludes the proof of Proposition \ref{prop:k-user}.

\subsection{Extension to a Transmitter with Multiple Antennas}
\label{subsec:miso}

As with B-DPC, S-SC can be extended to BCs with SIMO links, yielding Corollary \ref{cor:simo} again. Furthermore, S-SC can also be extended to BCs with multiple-input, single-output (MISO) links. In contrast, it is unclear how to accomplish this with B-DPC, because DPC is generally suboptimal in multiple-input Gaussian channels without utilizing the channel gain vector for precoding \cite{caire03:it}.

Consider a $K$-user quasi-static fading Gaussian BC without transmit CSI, with the transmitter equipped with $m_\mathrm{t}$ antennas. The $k$th receiver output is given by
\begin{eqnarray}
\label{eqn:miso}
\rvy_k[n] = \sum_{m = 1}^{m_\mathrm{t}} \rvh_{k, m} \rvx_m[n] + \rvz_k[n],\quad n = 1, \ldots, N.
\end{eqnarray}
The i.i.d. additive noises $\rvz_k[\cdot] \sim \mathcal{CN}(0, 1)$. The vector inputs $\bigl[ \rvx_1[\cdot], \ldots, \rvx_{m_\mathrm{t}}[\cdot] \bigr]^\mathrm{T}$ have an average power constraint $\snr$, {\it i.e.},
\begin{eqnarray}
\frac{1}{N}\sum_{n = 1}^N \sum_{m = 1}^{m_\mathrm{t}} |\rvx_m[i]|^2 \leq \snr,
\end{eqnarray}
over the coding block of length $N$. The complex-valued random variable $\rvh_{k, m}$ denotes the fading coefficient for the link from the $m$th transmit antenna to the $k$th user. We have the following result.

\begin{cor}
\label{cor:miso}
For the channel model (\ref{eqn:miso}), S-SC achieves an $\underline{\epsilon}$-outage achievable rate region identical to that described by $\mathcal{R}^\ast(\snr, \underline{\epsilon})$ for the $K$-user quasi-static fading scalar Gaussian BC model (\ref{eqn:k-user}), with $\rva_k$ replaced by $(1/m_\mathrm{t})\sum_{m = 1}^{m_\mathrm{t}}|\rvh_{k, m}|^2$.
\end{cor}

{\it Case Study: Multiple Transmit Antennas of Spatially Uncorrelated Rayleigh Fading}

\begin{figure}[p]
\psfrag{xlabel}{$R_1$ (nats)}
\psfrag{ylabel}{$R_2$ (nats)}
\epsfxsize=5in
\epsfclipon
\centerline{\epsffile{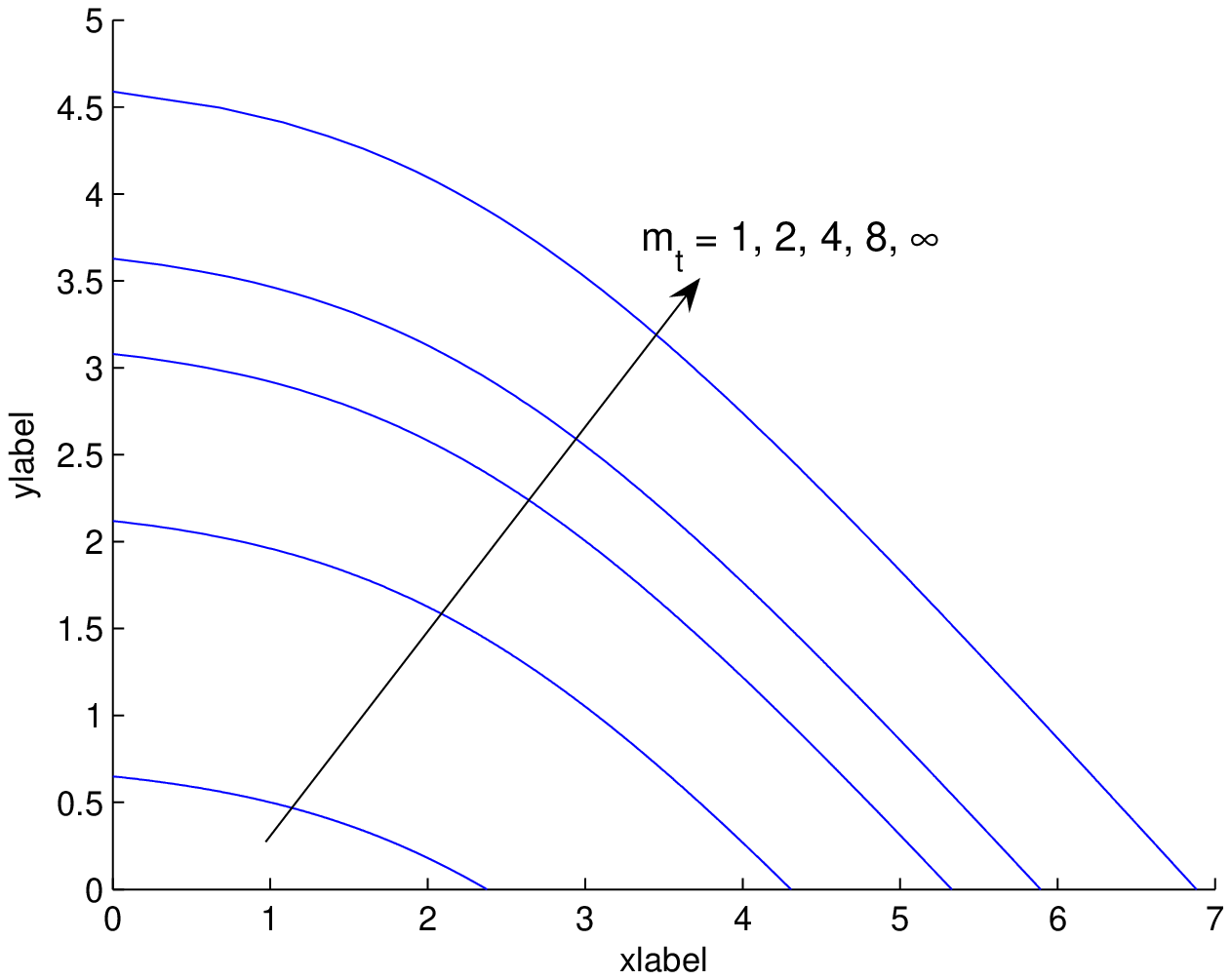}}
\caption{Outage achievable rate regions for a two-user BC with multiple transmit antennas experiencing spatially uncorrelated Rayleigh fading.}
\label{fig:miso}
\end{figure}

Unlike receivers in a typical downlink system, the physical size of the transmitter is usually less constrained. Consequently, multiple transmit antennas without spatial correlation may be deployed. Here we examine the case of two receivers each equipped with a single antenna, and with each link experiencing Rayleigh fading independent of the others. We assume that the two receivers are under a near-far situation, with the mean of each fading coefficient of the first receiver being $10$ and that of the second being $1$. The target outage probability vector is $\underline{\epsilon} = [0.01 \; 0.01]$, and the average power constraint $\rho$ is $20$dB. Figure \ref{fig:miso} depicts the $\underline{\epsilon}$-outage achievable rate regions $\mathcal{R}^\ast(\rho, \underline{\epsilon})$, for different values of the number of transmit antennas $m_\mathrm{t}$. It is clearly illustrated that multiple transmit antennas substantially enlarge the outage achievable rate region.

\section{Concluding Remarks}
\label{sec:conclude}

In this paper, we consider downlink transmission modeled as a quasi-static fading Gaussian BC without transmit CSI. We identify a non-trivial outage achievable rate region which always dominates that of time-sharing. We show that there exist two distinct coding schemes, namely B-DPC and S-SC, both achieving this outage achievable rate region. The analysis of these coding schemes highlights the statistical nature of the communication problem under an outage criterion. That is, in order to be outage-efficient, it is not the performance for individual channel realizations, but instead the performance statistics, that play a key role.

\section*{Acknowledgment}
The authors wish to thank Giuseppe Caire for encouragement and useful comments in preparing this paper.

\bibliographystyle{ieee}
\bibliography{v095}

\begin{thebibliography}{1}

\bibitem{cover72:it}
T.~M. Cover,
\newblock ``Broadcast Channels,''
\newblock {\em \textit{IEEE Trans. Inform. Theory}}, vol. 18, no. 1, pp. 2--14, Jan. 1972.

\bibitem{cover98:it}
T.~M. Cover,
\newblock ``Comments on Broadcast Channels,''
\newblock {\em IEEE Trans. Inform. Theory}, vol. 44, no. 6, pp. 2524--2530, Oct. 1998.

\bibitem{bergmans74:it-2}
P.~P. Bergmans and T.~M. Cover,
\newblock ``Cooperative Broadcasting,''
\newblock  {\em \textit{{IEEE} Trans. Inform. Theory}}, vol. 20, no. 3, pp. 317--324, May 1974.

\bibitem{bergmans74:it-1}
P.~P. Bergmans,
\newblock ``A Simple Converse for Broadcast Channels with Additive White Gaussian Noise,''
\newblock {\em \textit{IEEE Trans. Inform. Theory}}, vol. 20, no. 2, pp. 279--280, Mar. 1974.

\bibitem{costa83:it}
M.~H.~M. Costa,
\newblock ``{Writing on Dirty Paper},''
\newblock {\em \textit{{IEEE} Trans. Inform. Theory}}, vol. 29, no. 3, pp. 439--441, May 1983.

\bibitem{caire03:it}
G. Caire and S. Shamai (Shitz),
\newblock ``{On the Achievable Throughput of a Multi-antenna Gaussian Broadcast Channel},''
\newblock {\em \textit{{IEEE} Trans. Inform. Theory}}, vol. 49, no. 7, pp. 1691--1706, Jul. 2003.

\bibitem{yu04:it}
W. Yu and J. M. Cioffi,
\newblock ``Sum Capacity of Gaussian Vector Broadcast Channels,''
\newblock {\em \textit{{IEEE} Trans. Inform. Theory}}, vol. 50, no. 9, pp. 1875--1892, Sep. 2004.

\bibitem{viswanath02:dimacs}
P. Viswanath and D. N. C. Tse,
\newblock ``On the Capacity of the Multiple Antenna Broadcast Channel,''
\newblock {\em \textit{DIMACS Workshop on Signal Processing for Wireless Communications}}, Oct. 2002.

\bibitem{vishwanath03:it}
S. Vishwanath, N. Jindal, and A. Goldsmith,
\newblock ``Duality, Achievable Rates, and Sum-Rate Capacity of Gaussian MIMO Broadcast Channels,''
\newblock {\em \textit{{IEEE} Trans. Inform. Theory}}, vol. 49, no. 10, pp. 2658--2668, Oct. 2003.

\bibitem{weingarten06:it}
H. Weingarten, Y. Steinberg, and S. Shamai (Shitz),
\newblock ``The Capacity Region of the Gaussian Multiple-Input Multiple-Output Broadcast Channel,''
\newblock {\em \textit{{IEEE} Trans. Inform. Theory}}, vol. 52, no. 9, pp. 3936--3964, Sep. 2006.

\bibitem{li01:it-1}
L. Li and A.~J. Goldsmith,
\newblock ``Capacity and Optimal Research Allocation for Fading Broadcast Channels, -- Part I: Ergodic Capacity,''
\newblock {\em IEEE Trans. Inform. Theory}, vol. 47, no. 3, pp. 1083--1102, Mar. 2001.

\bibitem{li01:it-2}
L. Li and A.~J. Goldsmith,
\newblock ``{Capacity and Optimal Resource Allocation for Fading Broadcast Channels, -- Part II: Outage Capacity},''
\newblock {\em \textit{{IEEE} Trans. Inform. Theory}}, vol. 47, no. 3, pp. 1103--1127, Mar. 2001.

\bibitem{tuninetti03:isit}
D. Tuninetti and S. Shamai (Shitz),
\newblock ``The Capacity Region of Two User Fading Broadcast Channels with Perfect Channel State Information at the Receivers,''
\newblock {\em in Proc. IEEE Int. Symp. Inform. Theory (ISIT)}, Yokohama, Japan, 2003.

\bibitem{han03:book}
T.~S. Han,
\newblock {\em Information-Spectrum Methods in Information Theory},
\newblock Springer-Verlag, Berlin, 2003.

\bibitem{iwata05:ieice}
K-i. Iwata and Y. Oohama,
\newblock ``Information-Spectrum Characterization of Broadcast Channel with General Source,''
\newblock {\em \textit{IEICE Trans. Fundamentals}}, vol. E88-A, no. 10, pp. 2808--2818, Oct. 2005.



%
%

\bibitem{gelfand80:pcit}
S.~I. Gel'fand and M.~S. Pinsker,
\newblock ``Coding for Channel with Random Parameters,''
\newblock {\em \textit{Problems of Control and Information Theory}}, vol. 9, no. 1, pp. 19--31, 1980.

\end{thebibliography}

\end{document}